\documentclass{JHEP}
\usepackage{epsfig,rotate,array,amsmath,amsfonts,amssymb,rotating}
\newcommand{\beq}{\begin{equation}}
\newcommand{\eeq}{\end{equation}}

\newcommand{\Z}{{\mathbb Z}}
\newcommand{\C}{{\mathbb C}}

\def\omit#1{}

\title{Dimer models and toric diagrams}

\author{Amihay Hanany$^1$, Kristian D. Kennaway$^2$\\
~\\
$^1$ Center for Theoretical Physics,\\
Massachusetts Institute of Technology,\\
Cambridge, MA 02139, USA.\\
~\\
$^2$ Department of Physics,\\
University of Toronto,\\
Toronto M5S 1A7, Ontario, Canada.\\
}

\abstract{We propose a duality between quiver gauge theories and the combinatorics of dimer models.  The connection is via toric diagrams together with multiplicities associated to points in the diagram (which count multiplicities of fields in the linear sigma model construction of the toric space).  These multiplicities may be computed from both sides and are found to agree in all known examples.  The dimer models provide new insights into the quiver gauge theories: for example they provide a closed formula for the multiplicities of arbitrary orbifolds of a toric space, and allow a new algorithmic method for exploring the phase structure of the quiver gauge theory.}


\preprint{MIT-CTP-3613\\
hep-th/0503149}

\begin{document}
\section{Introduction}

Dimer models\footnote{{\it dimer} (n): a compound whose molecules are composed of two identical monomers.} are two-dimensional statistical mechanical models of tilings of graphs.  They have been recently introduced to the string theory literature in \cite{Okounkov:2003sp},
%
which proposed to relate the dimer models to the target space topological string field theory.  This viewpoint has been developed \cite{Iqbal:2003ds} for the crystal models also discussed in \cite{Okounkov:2003sp}, but the precise topological string theory interpretation of the dimers has not yet been made clear\footnote{See however the recent papers \cite{Stienstra:2005a,Stienstra:2005b} which pointed out that the Ronkin function, which is associated to the limit curve of the dimer model, can be used to recover the instanton contributions to the topological string partition function.}.

In this paper we develop a different string theoretical interpretation of the dimer models, and relate them to the gauge theories that live on the world-volume of D-branes probing non-compact toric spaces. These theories are generally known under the name of quiver gauge theories and were the subject of study of many recent papers. A generic feature of these theories is that for a given singular non-compact manifold there exist infinitely many quiver gauge theories that probe the same singularity. These theories are all related by Seiberg duality and form a duality tree that encodes the relations between theories. Examples of such trees and a discussion of this phenomenon can be found in \cite{Franco:2003ea}. Out of all the theories in the duality tree there is a special subclass of theories which are called the toric phases of the quiver gauge theory. In all known examples this subclass contains a finite number of theories. The toric phases are characterized by having the same rank for all gauge groups in the quiver, with each field appearing exactly twice in the superpotential. A discussion of all of the toric phases for a class of quiver gauge theories called $Y^{p,q}$ was recently performed in \cite{Benvenuti:2004wx}.

This paper is based on the observation that the coefficients appearing in the determinant of the Kasteleyn matrix of the graph associated to a dimer model are the same numbers that count the multiplicities of fields in the linear sigma model realization of the toric variety.  Moreover, the various toric phases of the quiver gauge theories, which are related by toric duality, may all be obtained from related dimer models.  Thus, the dimer models know about the phase structure of the gauge theories living on the moduli space of D-branes probing toric singularities.  This paper explores some aspects of this correspondence.

In section \ref{sec:quivers} we revise some basics of quiver gauge theories, and the forward and inverse algorithms for going between the quiver theory and its' moduli space.  In section \ref{sec:dimers} the dimer models are introduced and their relation to toric geometry is elucidated.  In terms of the dimer models, certain aspects of the toric geometry become clear: for example they provide simple formul\ae\ for the linear sigma model field multiplicities of orbifolds of any non-compact toric 3-fold, which recover the known examples in particular cases.

We show how the operation of Higgsing the gauge theory to remove points from the toric diagram\footnote{also called ``partial resolution'' in previous string theory literature.} corresponds to cutting edges of the graph, and describe how this procedure may be used to systematically enumerate the toric phases of a given singularity.

In a subsequent paper we will explore the map between the quiver gauge theory and dimer pictures and combine it with the results from the dimer side to obtain a deeper understanding of the quiver gauge theories and their embedding in string theory.

\section{Review of quivers and toric singularities}
\label{sec:quivers}

When D-branes sit on a non-compact singular Calabi-Yau (CY) manifold, $X$, the world volume theory living on them is a quiver gauge theory which involves gauge fields, matter fields and interaction terms in a theory that is generically supersymmetric with 4 supercharges. In special cases when the manifold has more symmetries, the supersymmetry may be enhanced to as many as 16 supercharges. Taking the number of D-branes to be $N$, examples are $X=\C^3$ with quiver gauge theory SYM $U(N)$ and 16 supercharges; $X=\C\times\C^2/\Z_n$ with quiver gauge theory having 8 supercharges and specified by the affine Dynkin diagram of ${\hat{A}}_{n-1}$, etc.

The problem of finding a quiver gauge theory for a given toric singularity is a difficult problem and is still not fully understood. One of the methods in constructing this quiver gauge theory is known under the name ``partial resolution of Abelian orbifolds" \cite{Morrison:1998cs,Beasley:1999uz}. An algorithm for computing the gauge theory for a given toric singularity was used in \cite{Feng:2000mi} to find the gauge theories corresponding to all of the toric del Pezzo surfaces.\footnote{In the literature the process of finding the toric data for a given quiver gauge theory is called the ``Forward Algorithm" and the process of finding quiver gauge theories from a given toric data is called the ``Inverse Algorithm".}

The sketch of the algorithm goes as follows. To every toric diagram one can embed it in a sufficiently large toric diagram with a known quiver gauge theory. Such a diagram can be chosen to be $\C^3/\Z_n\times\Z_m$ with $n$ and $m$ the smallest integers such that the orbifold toric diagram (which looks like a triangle with sides of dimension $(n+1)\times(m+1)$), contains the toric diagram of interest. Then one can remove points by introducing FI terms to the corresponding gauge theories. This leads to a Higgs mechanism and generation of mass terms for some of the fields, which can then be integrated out.  As we follow through the process of removing points one can end up with the right quiver and superpotential for the desired toric diagram. While this is a simple process to describe, it becomes impractical to implement for large toric diagrams. With dimers, however, the amount of computation reduces substantially and it is possible to proceed with this algorithm for larger toric diagrams. See section \ref{partial_resolution} for more details.

One interesting aspect of the algorithm is its non-uniqueness in producing a quiver gauge theory, thus leading to a one-to-many map between a geometric singularity and quiver gauge theories. The corresponding quiver gauge theories are different UV theories but flow to the same universality class in the IR, characterized by the singular geometry.
The non-uniqueness phenomenon is called ``Toric duality" \cite{Feng:2000mi} and is found to be equivalent to Seiberg duality \cite{Beasley:2001zp,Feng:2001bn,Cachazo:2001sg}.
Each different quiver theory resulting from the algorithm is called a ``toric phase" and has the property that all ranks of the gauge groups are equal to the number of D-branes, $N$, and every bi-fundamental field appears in the superpotential precisely twice: once with a positive sign and once with a negative sign.

The computation of the gauge groups and matter fields turns out to be a simpler task than that of computing the superpotential. Indeed a few methods, different and even faster than the inverse algorithm of \cite{Feng:2000mi}, were used in the literature to compute the matter content of the quiver theory: using (p,q) webs \cite{Hanany:2001py}, Ext groups \cite{Cachazo:2001sg}, exceptional collections \cite{Herzog:2003dj}, and others. Recently a generic computation of superpotentials was proposed in \cite{Aspinwall:2004bs}. One of the results of the present work in relating dimers to quiver gauge theories is a major simplification in computing superpotentials. If, using independent data, one comes up with a guess for a quiver gauge theory, then one of the immediate subsequent computations to be performed is the construction of its' moduli space of vacua. This moduli space coincides with the CY manifold probed by the D-branes. The dimer methods that we introduce in this paper allow a quick computation of the toric data of this CY manifold, much faster than the forward algorithm or any other existing methods do.

Another interesting aspect of the algorithm used in \cite{Feng:2000mi} is the appearance of a {\bf multiplicity} of fields in the gauged linear sigma model. A detailed study of these multiplicities and of toric duality, done in \cite{Feng:2002zw}, proposes that the different toric phases of a given singular manifold are characterized by different multiplicities. All known examples studied so far have the property that the multiplicities of the toric diagram are different for different toric phases. Using the methods presented in this paper it becomes clear that there are counterexamples to this proposal as the results on $Y^{6,0}$ indicate. Furthermore these multiplicities have beautiful combinatorial properties and the explicit formulas given in this paper shed light on their origin.

The ``Forward Algorithm" starts with a quiver gauge theory for one D-brane, ($N=1$). In the language of 3+1 dimensional supersymmetry with 4 supercharges it has $n$ $U(1)$ vector multiplets, $m$ bi-fundamental chiral multiplets and $m-n$ terms in the superpotential. At the end of the computation one ends up with $c$ fields $p_i$ \footnote{In the math literature the fields $p_i$ are the homogeneous coordinates of the toric variety. See \cite{Muto:2001gu} for a detailed discussion on the $\C^3/\Z_2\times\Z_2$ orbifold.} subject to $c-3$ moment map relations. This gives a $(c-3)\times c$ charge matrix which is called $Q_t$. The integral co-kernel of $Q_t$ is a $3\times c$ matrix called $G_t$ with columns corresponding (up to repetition) to nodes of the three dimensional toric diagram. All these nodes lie on a 2 dimensional sub-lattice, due to the CY condition. Due to the repetition, each node comes with a multiplicity and it is these multiplicities that  form the main interest of the present paper.  
Further details on the forward algorithm can be found in \cite{Feng:2000mi,Feng:2001xr} and in \cite{Dunn:2005a}.

The forward algorithm uses linear algebra and integer programing and turns out to be a useful tool in computing toric data, including their multiplicities, for sufficiently small quivers. Here small means small $n$ and small $m$. Once the size of $m$ and $n$ become large it becomes exponentially difficult to compute the multiplicities and hence practically impossible. For example the computation of the multiplicities for the quiver theory of $\C^3/\Z_4\times\Z_4$ which has $n=16$ and $m=48$ would take the order of few days on an average-speed desktop computer. Obviously it becomes clear that we need more efficient methods to compute these multiplicities for generic toric diagrams. As we will see in this paper, dimers provide the right tool to compute these multiplicities with arbitrary length and with analytic expressions. In particular for orbifolds of $\C^3$, of the conifold, or of any other type of singularity, we have explicit expressions which were found by Kenyon \cite{Kenyon:2003uj} and are given in equation \eqref{eq:orbifold} and generalized in \eqref{oneorbifold}.

The collection of multiplicities which were known so far were summarized in \cite{Feng:2002zw}. Some others were found in \cite{Sarkar:2000iz}.  A few more, found using the forward algorithm, will be published in an upcoming work \cite{Dunn:2005a}. One of the points of this paper is the ability to generalize all these results into a simple computational tool which produces many more new results.

In a recent paper \cite{Benvenuti:2004dy} the quiver gauge theories of the $Y^{p,q}$ manifolds were found. A subsequent work \cite{Benvenuti:2004wx} computed all the toric phases for each $Y^{p,q}$ model and demonstrated the rich structure they have. One non-trivial check for the duality between the $Y^{p,q}$ manifolds and their quiver gauge theories was performed in computing volumes of supersymmetric cycles and comparing them to $R$-charges computed in the gauge theory. Another check which follows from this work is to perform the computation of the moduli space of vacua of the gauge theories using their dimer description.
A subsequent work \cite{Brian} will present another infinite family of quiver gauge theories which have no known metrics. The methods in the current paper will provide a consistency check for the toric description of the corresponding CY manifolds.

The geometrical meaning of these multiplicities still remains unclear, and it remains unknown whether they are connected in any way to topological strings.  We expect that studying  such connections will be a highly fruitful area of research.

This work introduces a connection between dimers and multiplicities of toric diagrams, relevant to quiver gauge theories. In a subsequent publication \cite{Kennaway:2005b} we will explore the relationship between dimers and quiver gauge theories in greater detail, by extending the correspondence between quiver gauge theories and toric geometries as their moduli space. Dimers turn out to be the right tool to study both in a simple and efficient fashion.

\section{Dimer models} 
\label{sec:dimers}

This section is based on \cite{Kenyon:2003uj, Kenyon:2002a}.

A {\bf bipartite graph} is a graph with the property that all vertices can be coloured black or white, such that every black vertex is only adjacent to white vertices, and vice versa. A {\bf perfect matching} of a bipartite graph is a subset of edges (``dimers'') such that every vertex in the graph is an endpoint of precisely one such edge.  A {\bf dimer model} is the statistical mechanics of such a system, i.e.~of random perfect matchings of the graph with assigned edge weights.  We will be interested in dimers on doubly-periodic graphs, i.e.~defined on the torus $T^2$.

Many important properties of the dimer model are governed by the {\bf Kasteleyn matrix} $K(z,w)$, a weighted, signed adjacency matrix of the graph with (in our conventions) the rows indexed by the white nodes, and the columns indexed by the black nodes.  It is constructed as follows:

To each edge in the graph, multiply the edge weight by $\pm 1$ so that around every face of the graph the product of the edge weights over edges bounding the face has the following sign

\begin{equation}
\mbox{sign}(\prod_i e_i) = \left\{\begin{array}{cc}
+1 &  {\mbox \it if} \ (\mbox{\# edges}) = 2 \mod 4 \\
-1 & {\mbox \it if} \ (\mbox{\# edges}) = 0 \mod 4
\end{array}
\right.
\label{eq:signs}
\end{equation}
It is always possible to arrange this \cite{Kasteleyn}.

The colouring of vertices in the graph induces an orientation to the edges, for example the orientation ``black'' to ``white''.  Construct paths $\gamma_w$, $\gamma_z$ in the dual graph that wind once around the $(0,1)$ and $(1,0)$ cycles of the torus, respectively.  For every edge crossed by $\gamma$, multiply the edge weight by a factor of $w$ or $1/w$ (respectively $z$, $1/z$) according to the relative orientation of the edges in $G$ crossed by $\gamma$.

The adjacency matrix of the graph $\Gamma$ weighted by the above factors is the Kasteleyn matrix $K(z,w)$ of the graph.  The determinant of this matrix $P(z,w) = \det K$ is called the characteristic polynomial of the graph, and will be our main object of study.

Given an arbitrary reference matching $M_0$ on the graph, the difference $M - M_0$ defines a set of closed curves on the graph in $T^2$.  This in turn defines a height function on the faces of the graph: when a path in the dual graph crosses the curve, the height is increased or decreased by 1 according to the orientation of the crossing.  A different choice of reference matching $M_0$ shifts the height function by a constant.  Thus, only differences in height are physically significant.

In terms of the height function, the characteristic polynomial takes the following form:

\begin{equation}
\label{eq:det}
P(z,w) = z^{h_{x0}} w^{h_{y0}} \sum c_{h_x,h_y} (-1)^{h_x + h_y + h_x h_y} z^{h_x} w^{h_y}
\end{equation}
where $c_{h_x,h_y}$ are integer coefficients that count the number of paths on the graph with height change $(h_x,h_y)$ around the two fundamental cycles of the torus.  These coefficients do not have an obvious geometrical meaning in toric geometry\footnote{There should be an interpretation in terms of the symplectic quotient construction of the space.}, but will turn out to be equal to the multiplicity of fields required to engineer the toric variety as the moduli space of a D-brane world-volume quiver gauge theory.

In the following it will be convenient to define the following quantity:
\begin{equation}
\label{ctot}
c=\sum |c_{h_x,h_y}|,
\end{equation}
which counts the total multiplicity of the linear sigma model fields. This number was computed for special cases \cite{Sarkar:2000iz,Feng:2002zw,Muto:2002pk}.

The overall normalization of $P(z,w)$ is not physically meaningful: since the graph does not come with a prescribed embedding into the torus (only a choice of periodicity), the paths $\gamma_{z,w}$ winding around the primitive cycles of the torus may be taken to cross any edges en route.  Different choices of paths $\gamma$ multiply the characteristic polynomial by an overall power $z^i w^j$, and by an appropriate choice of path $P(z,w)$ can always be normalized to contain only non-negative powers of $z$ and $w$.

\EPSFIGURE{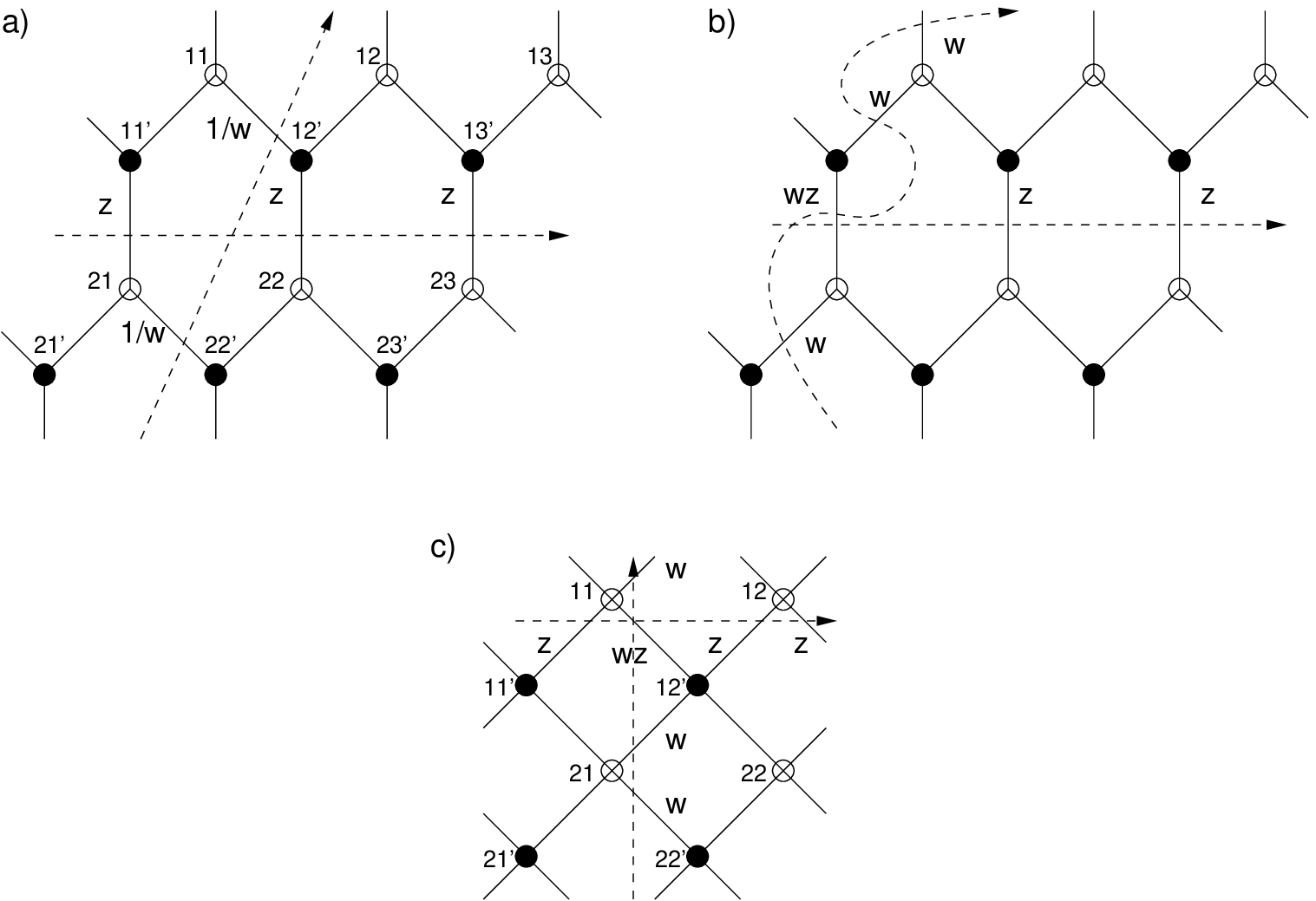}{The graphs corresponding to orbifolds a) $\C^3/\Z_3 \times \Z_2$, b) the same graph with a different choice of path $\gamma_w$, c) ${\cal C}/\Z_2 \times \Z_2$ an orbifold of the conifold\label{fig:orbifolds}}

For example, figure \ref{fig:orbifolds}a produces the Kasteleyn matrix and determinant

\begin{eqnarray}
K(z,w) &=& \left(\begin{array}{cccccc}
1&-1/w&0&-1&0&0\\
0&1&-1&0&-1&0\\
-1&0&1&0&0&-1\\
-z&0&0&1&-1/w&0\\
0&-z&0&0&1&-1\\
0&0&-z&-1&0&1
\end{array}\right)\nonumber \\
P(z,w) &=& \frac{1}{w^2} - \frac{2}{w} + 1 - 3 z - \frac{6 z}{w} + 3 z^2 - z^3 \equiv P_0
\end{eqnarray}
while the choice of path in figure \ref{fig:orbifolds}b (which is chosen so that all crossings occur with positive orientation) gives
\begin{eqnarray}
K(z,w) &=& \left(\begin{array}{cccccc}
w&-1&0&-w&0&0\\
0&1&-1&0&-1&0\\
-1&0&1&0&0&-1\\
-z w&0&0&w&-1&0\\
0&-z&0&0&1&-1\\
0&0&-z&-1&0&1
\end{array}\right)\nonumber \\
P(z,w) &=& 1 - 2w + w^2 -  3 z w^2 - 6 z w + 3 z^2 w^2 - z^3 w^2 \nonumber\\
&=& w^2 P_0
\end{eqnarray}

The degrees of the monomials of $P$ define points in the lattice $\Z^2$.  The convex hull of these points is the Newton polygon of the graph, which provides the link between dimer models and toric geometry.  

For a toric three-fold the set of torus actions

\begin{equation}
(z_1, z_2, z_3) \mapsto (\lambda^{n_1} z_1,\lambda^{n_2} z_2,\lambda^{n_3} z_3)
\end{equation}
where $\lambda \in \C^*$, defines a set of lattice vectors $\{(n_1, n_2, n_3) \in \Z^3\}$ which generate a fan $G$ of convex cones.  The Calabi-Yau condition is equivalent to the statement that the lattice vectors are all coplanar, and their projection onto this hyperplane defines a convex polygon in $\Z^2$, the Newton polygon.  

The generators of the dual cone $\check G$ define the toric variety algebraically by

\begin{equation}
V = \mbox{Spec}(\C[X_i^{\check G \cap \Z^3}])
\end{equation}
i.e.~in terms of the algebraic equations satisfied by the monomials $X_1^a X_2^b X_3^c$ associated to elements $(a,b,c)$ in the dual fan.  They correspond to the degeneration locus (``toric skeleton'') of the $U(1)^n$ torus actions defining the toric variety in the symplectic quotient (linear sigma model) construction of $V$.  In string theory language the generators of the dual fan (after projection) are the charges $(p,q)$ of 5-branes in the dual brane-web construction of the toric variety. See \cite{Aharony:1997bh} for further details on $(p,q)$ 5-brane webs.

The action of $SL(2,\Z)$ on the torus (i.e.~on the periodicity vectors) induces an action of $SL(2,\Z)$ on the Newton polygon as a subset of $\Z^2$.  This $SL(2,\Z)$ action is also visible in the $(p,q)$ 5-brane web, where it acts on the charge vector.

\subsection{Examples}

It is a simple exercise to write down some periodic bipartite graphs and compute their Kasteleyn determinants as described above.  Figure \ref{fig:graph1} shows the important examples of the hexagonal and square graphs (which will be of particular importance to us), and the square-octagon graph. Incidentally, these are three of the four different periodic plane tilings with even equilateral polygons\footnote{We thank R. Kenyon for a correspondence on this point.}.

\EPSFIGURE{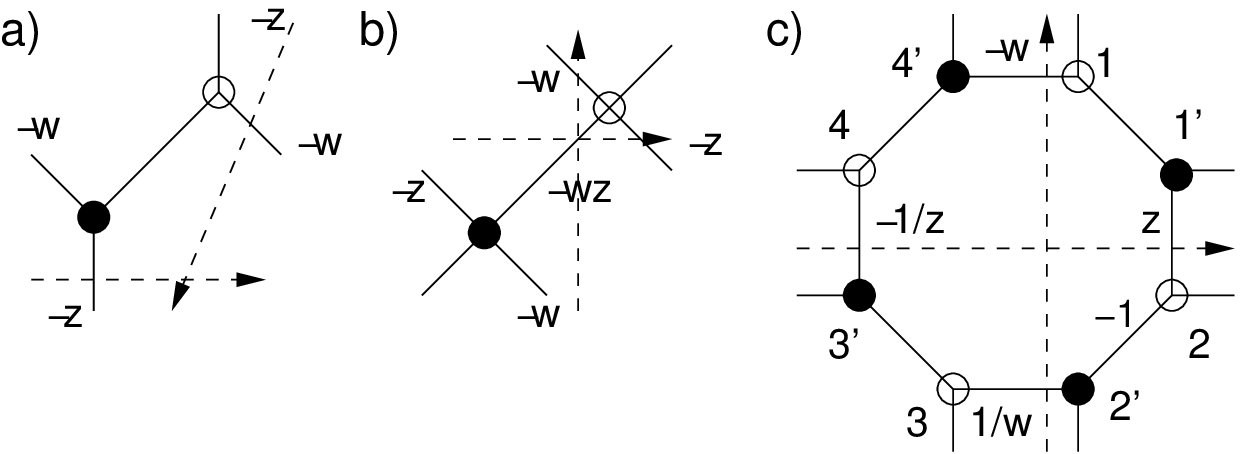}{The fundamental domain of a) the hexagonal graph corresponding to $\C^3$, b) the square graph corresponding to the conifold ${\cal C}$, and c) the square-octagon graph corresponding a toric phase of the cone over the Hirzebruch surface $F_0$.  The dashed lines show the periodicity of the graph, i.e.~how neighbouring domains adjoin.  Also marked are an example choice of signs and weights associated to the construction of Kasteleyn matrix.\label{fig:graph1}}

For the hexagonal graph (which will correspond to $\C^3$) the Kasteleyn matrix is $1 \times 1$, and with the choice of signs and paths $\gamma$ indicated in figure \ref{fig:graph1}a we find

\begin{equation}
P(z,w) = \det K(z,w) = 1 - z - w
\end{equation}
The corresponding Newton polygon is also shown in figure \ref{fig:braneweb}a together with the absolute value of the coefficients.

\EPSFIGURE{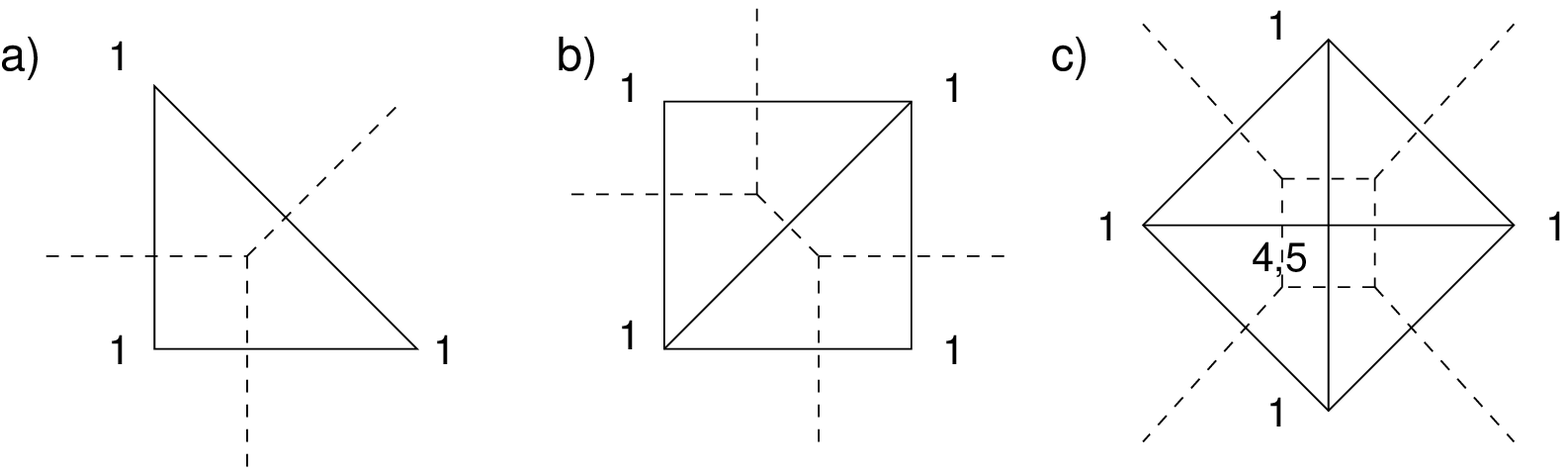}{The Newton polygon (solid lines)and dual 5-brane web (dashed lines), and the linear sigma model field multiplicities for a) $\C^3$, b) the (resolved) conifold, c) the two toric phases of the cone over $F_0$.\label{fig:braneweb}}

Similarly, for the square graph we find 

\begin{equation}
P(z,w) = \det K(z,w) = 1 - z - w - z w
\end{equation}
which agrees with the Newton polygon of the conifold ${\cal C}$.

For the square-octagon graph the Kasteleyn matrix is

\begin{eqnarray}
K(z,w) &=& \left(\begin{array}{cccc}
1&1&0&-1/w\\
1/z&-1&1&0\\
0&w&1&1\\
1&0&-z&1\\
\end{array}
\right)\nonumber\\
P(z,w) &=& \det K(z,w) = -1/w - w - 1/z - z - 5
\label{eq:phase5}
\end{eqnarray}
as shown in figure \ref{fig:braneweb}c.  The polygon is equal to the toric diagram of the cone over the Hirzebruch surface $F_0$, and the multiplicity `5' of the internal point agrees with one of the two toric phases of this surface, computed in \cite{Feng:2002zw} (we will explain in the next section how to obtain the other phase). This phase is called in the language of \cite{Benvenuti:2004wx} $Y^{2,0}$ with one double impurity.

\subsection{Orbifolds}
\label{sec:orbifolds}

Given the fundamental domain of a graph, a natural operation is to enlarge the domain of the graph by taking $m \times n$ copies of the fundamental domain.  It is easily verified that this corresponds to subdividing the lattice of the toric diagram, i.e.~to orbifolding the space by $\Z_m \times \Z_n$, where the orbifold action is generated by

\begin{eqnarray}
(z_1, z_2, z_3) \mapsto (\lambda z_1, z_2, \lambda^{-1} z_3),\hspace{0.2in} \lambda^n = 1 \\ 
(z_1, z_2, z_3) \mapsto (z_1, \omega z_2, \omega^{-1} z_3),\hspace{0.2in} \omega^m = 1
\end{eqnarray}

However, the multiplicities associated to points in the subdivided diagram will also change under orbifolding, and it was previously not known how to compute them in terms of the un-orbifolded quantities.  This problem is solved by using the Kasteleyn matrix of the dimer model.  The characteristic polynomial of the graph obtained by taking $n \times m$ copies of a graph is given by \cite{Kenyon:2003uj}

\begin{equation}
P_{n,m}(z,w) = \prod_{i=0}^{n-1} \prod_{j=0}^{m-1} P(\lambda^i z^{1/n},\omega^j w^{1/m})
\label{eq:orbifold}
\end{equation}
where $\lambda^n = 1, \omega^m=1$.

Notice that in order to use this formula one does not need to know the structure of the graph, only its' characteristic polynomial, i.e.~toric data and multiplicities.  It can thus be applied immediately to compute the multiplicities associated to {\it any} orbifold of {\it any} non-compact toric Calabi-Yau for which the multiplicities are known.  However in cases where the adjacency matrix can be written down for the orbifold in generality (such as for $\C^3$ or the conifold, see section \ref{sec:block}), it is computationally easier to simply take the determinant of this matrix).

Here are some examples that may be easily calculated using this formula.  They agree with examples calculated in \cite{Feng:2000mi,Feng:2001xr,Feng:2002zw}, who also observed the pattern of binomial coefficients along the edges of the polygon.  This structure is now apparent from formula (\ref{eq:orbifold}), where it arises from the expansion of the product.

\begin{eqnarray}
\C^3/(\Z_3 \times \Z_3) : \left[\begin{array}{cccc}
1 & -3 & 3 & -1 \\
-3 & -21 & -3 &\\
3 & -3 & &\\
-1 & & &
\end{array}\right] &\C^3/(\Z_4 \times \Z_4) :& \left[\begin{array}{ccccc}
1 & -4& 6 & -4 & 1 \\
-4 & -124 & -124 &-4&\\
6 & -124 & 6 &&\\
-4 & -4 && & \\
1 & & & &
\end{array}\right]\nonumber\\
\C^3/(\Z_2 \times \Z_4) : \left[\begin{array}{ccc}
1 & -2 & 1 \\
-4& -12 &\\
6 & -2 &\\
-4 & & \\
1 & &
\end{array}\right]&\C^3/(\Z_5 \times \Z_5) :& \left[\begin{array}{cccccc}
1&-5&10&-10&5&-1\\
-5&-605&-1905&-605&-5&\\
10&-1905&1905&-10&&\\
-10&-605&-10&&&\\
5&-5&&&&\\
-1&&&&&
\end{array}\right]\nonumber\\
{\cal C}/(\Z_1\times \Z_2) : \left[\begin{array}{cc}
1 & -1 \\
-2& -2\\
1 & -1
\end{array}\right]&{\cal C}/(\Z_3 \times \Z_3) : &\left[\begin{array}{cccc}
1&-3&3&-1\\
-3&-105&-105&-3\\
3&-105&105&-3\\
-1&-3&-3&-1
\end{array}\right]\nonumber
\label{eq:orbifoldex}
\end{eqnarray}

The above formula may be extended to the case of diagonal $\Z_n$ orbifolds by projection of the terms appearing in the double product.  Faithful actions of the diagonal $\Z_n$ are generated by actions of the form $(1,a,b=n-a-1)$ (up to permutation), and we find for $\C^3$

\begin{equation}
P_n(z,w) = \prod_{i=0}^{n-1} P(\lambda^{a i} w^{a/n} z^{(n-a)/n} , \lambda^{(a+1) i} w^{(a+1)/n} z^{(n-a-1)/n})
\label{oneorbifold}
\end{equation}
where the weights of $w$ and $z$ are related to the coordinates of the corner points that remain after orbifolding, see figure \ref{fig:diagorb}.

For example (suppressing minus signs):

\begin{itemize}
\item{The diagonal orbifold $\C^3/\Z^3$ with action $(1,1,1)$, also known as the cone over $dP_0$, which agrees with \cite{Feng:2000mi},
\begin{equation}
\left[\begin{array}{ccc}
1&&\\
&3&1\\
&1&
\end{array}\right]
\end{equation}}
\item{The orbifold $\C^3/\Z_8$ with action $(1,6,1)$
\begin{equation}
\left[\begin{array}{ccc}
1&0&0\\
0&0&0\\
0&0&0\\
0 & 2 & 0\\
0 & 16 & 0\\
0 & 20 & 0\\
0 & 8 & 1\\
0 & 1 & 0
\end{array}\right]
\end{equation}
which is the toric phase of $Y^{4,4}$ \cite{Dunn:2005a}. Here, and in the following, 0's indicate points on the lattice which are not included in the toric diagram.
}
\item{The orbifold $\C^3/\Z_7$ with action $(1,3,3)$ \cite{Muto:2002pk,Yang}
\begin{equation}
\left[\begin{array}{ccccc}
1&0&0&0&0\\
0&7&0&0&0\\
0 & 0&14 & 0 & 0\\
0 & 0 &0& 7 & 1\\
0 & 0 & 0 &1 & 0\\
\end{array}\right]
\end{equation}
}
\item{The other distinct $\C^3/\Z_7$ orbifold with action $(1,2,4)$ \cite{Yang}
\begin{equation}
\left[\begin{array}{cccccc}
1&0&0&0&0&0\\
0&0&7&0&0&0\\
0 & 0&0 &7& 7 & 1\\
0 & 0 &0&0 &1 & 0\\
\end{array}\right]
\end{equation}
}
\end{itemize}
The results here are consistent with the analytic formula that was computed for $c$, the total multiplicity of the toric diagrams \eqref{ctot}, for an orbifold $\C^3/\Z_n$ with an action $(1,1,-2)$, \cite{Muto:2002pk}. For given $n$ the multiplicity is given by the Fibonacci numbers:

\EPSFIGURE[t]{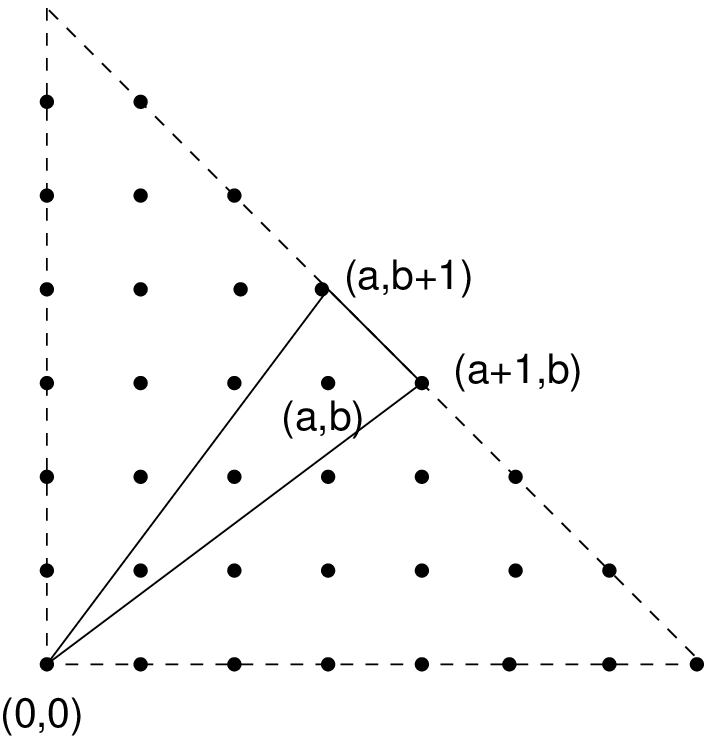,height=3in,width=3in}{The toric diagram of the diagonal orbifold of $\C^3/\Z_n$ with action $(1,a,b)$.\label{fig:diagorb}}

\begin{equation}
\label{fibonacci}
c=\left(\frac{1+\sqrt{5}}{2}\right)^n+\left(\frac{1-\sqrt{5}}{2}\right)^n+2.
\end{equation}

In terms of the graph, different orbifold actions correspond to different choices of periodicity of  adjacent copies of the $n \times m$ unit cell of the orbifold graph, i.e.~adjoining neighbouring unit cells with an offset\footnote{This looks very similar to the brane box constructions \cite{Hanany:1998it}.
Indeed, there is a deeper reason for this similarity; details will appear in the subsequent publication \cite{Kennaway:2005b}.}. For example, in figure \ref{fig:z2conifold} are shown two possible ways of adjoining neighbouring fundamental domains of the $\Z_2$ orbifold of the conifold.  These correspond to the two different actions of the $\Z_2$: figure \ref{fig:z2conifold}a corresponds to the action $(1,1,0,0)$ acting on the four fields $(w, x, y, z)$ of the conifold defined by the equation $w x = y z$, and figure \ref{fig:z2conifold}b gives the diagonal orbifold action by $(1,1,1,1)$.

\EPSFIGURE[b]{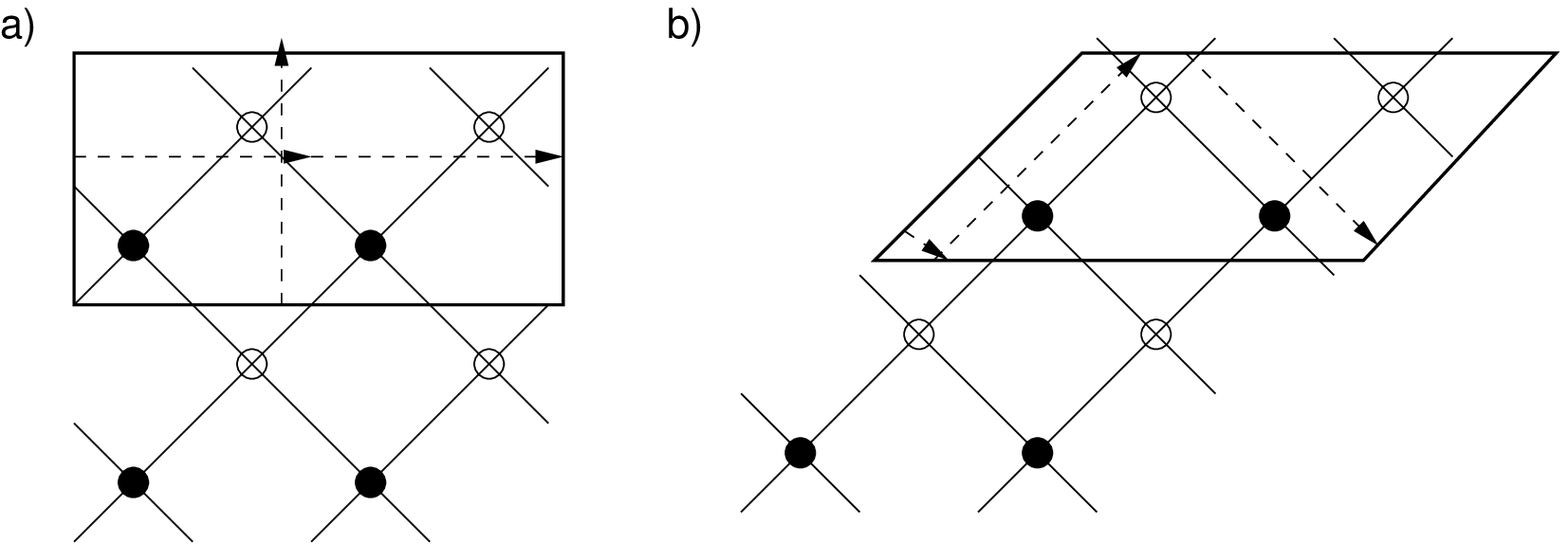}{The two orbifolds of the conifold by $\Z_2$ correspond to different ways adjacent copies of the $2 \times 1$ fundamental domain may be adjoined.\label{fig:z2conifold}}

Computing the Kasteleyn matrix we find for figure \ref{fig:z2conifold}a

\begin{eqnarray}
K(z,w) &=& \left(\begin{array}{cc}
z-1 & -w - w z\\
-1 - z & z - 1
\end{array}
\right)\\
P(z,w) &=& \det K(z,w) \nonumber\\
&=& 1 - w - 2z - 2 w z + z^2 - w z^2\nonumber
\end{eqnarray}
which agrees with the example computed already in (\ref{eq:orbifoldex}), and for  figure \ref{fig:z2conifold}b

\begin{eqnarray}
K(z,w) &=& \left(\begin{array}{cc}
-1+1/z & -1 - 1/w\\
-1-w & 1-z
\end{array}
\right)\nonumber\\
P(z,w) &=& \det K(z,w)\\
&=& -1/w - w + 1/z + z - 4\nonumber
\label{eq:phase4}
\end{eqnarray}
This phase in the language of \cite{Benvenuti:2004wx} is $Y^{2,0}$ with two single impurities and can indeed be derived as the $\Z_2$ orbifold of the conifold \cite{Franco:2005fd}. Comparing to (\ref{eq:phase5}) we see that this is another phase of the cone over $F_0$.  The two phases are related by Seiberg duality, as we will explain in followup work \cite{Kennaway:2005b}.

It is often the case that orbifolds of a space appear among the set of phases of larger spaces.  For example, taking the $\Z_1 \times \Z_3$ orbifold of phase `5' of $F_0$ gives the following multiplicities

\begin{equation}
\left[\begin{array}{ccccccc}
&&&-1&&&\\
-1&-15&-75&-140&-75&-15&-1\\
&&&-1&&&\\
\end{array}\right]
\end{equation}
and the corresponding orbifold of the phase '4' gives

\begin{equation}
\left[\begin{array}{ccccccc}
&&&-1&&&\\
-1&-12&-48&-76&-48&-12&-1\\
&&&-1&&&\\
\end{array}\right]
\end{equation}
which both appear among the 18 phases of $Y^{6,0}$, listed in appendix \ref{app:60}.

\section{Block determinant formul\ae\ for orbifolds}
\label{sec:block}

Since the action of orbifolding a space has a simple realization on the graph, it is often possible to write down the general form of the Kasteleyn matrix for arbitrary orbifolds, in the form of a block matrix.  The blocks are related to the adjacency matrix of the original fundamental domain (with a suitable choice of vertex labelling), and the arrangement of blocks corresponds to the interconnections between copies of the fundamental domain. Here we present the results for the orbifolds of $\C^3$ and of the conifold.  Generalizations to diagonal orbifold action, and to orbifolds of other spaces are also possible.

We make use of the block determinant formula for a block matrix $M$

\begin{equation}
M=\left(\begin{array}{cc}
A & B \\
C & D 
\end{array}\right)
\end{equation}
where $A, D$ are square matrices and $A$ is invertible.  Then

\begin{equation}
\det(M) = \det(A) \det(D-C A^{-1} B)
\label{eq:blockdet}
\end{equation}

\EPSFIGURE[t]{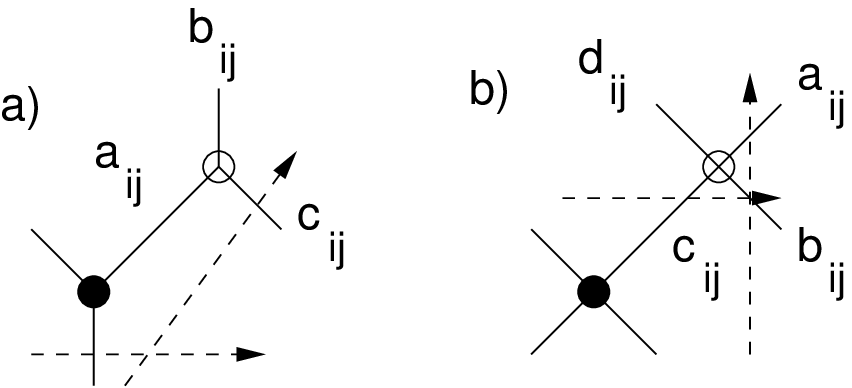}{The edge weights assigned to the $(i,j)$th copy of a fundamental domain.  The adjacency matrix for arbitrary multiples of the fundamental domain can be written down in block matrix form.\label{fig:edgeweights}}

\subsection{Orbifolds of $\C^3$}

With edge weights labelled according to figure \ref{fig:edgeweights}, it is easy to verify that the adjacency matrix of the orbifold $\C^3/(\Z_m \times \Z_n)$ takes the following $n m \times n m$ block matrix form:

\begin{equation}
K(z,w) = \left(\begin{array}{cccccc}
A_{1} &B_{1} w   & 0  &\cdots& \cdots & 0 \\
0 &A_{2}&    B_{2}& 0 &\cdots& 0 \\
\vdots&  \ddots   &\ddots&\ddots&\ddots & \vdots\\
0 & \ddots & \ddots & \ddots & A_{m-1} & B_{m-1} \\
B_{m} & 0 & \cdots & \cdots & 0 & A_{m}\\
\end{array} \right)
\end{equation}

where

\begin{eqnarray}
A_{i}&=&\left(\begin{array}{cccccc}
a_{i1} & b_{i1} & 0 & \cdots & \cdots & 0 \\
0 & a_{i 2} & b_{i 2} & 0 & \cdots & 0 \\
\vdots & \ddots & \ddots & \ddots & a_{i (n-1)} & b_{i(n-1)} \\
b_{i n} z & 0 &\cdots & \cdots&0 &a_{i n}
\end{array}\right)\nonumber\\
B_{i}&=&\left(\begin{array}{ccccc}
c_{i1} & 0 & \cdots & \cdots & 0 \\
0 & c_{i 2} & 0 & \cdots & 0 \\
\vdots & \ddots & \ddots & \ddots & \vdots \\
0 & \cdots & \cdots & 0 & c_{i n}\\
\end{array}\right)
\end{eqnarray}

Using the block determinant formula (\ref{eq:blockdet}) the evaluation of the determinant can be reduced to

\begin{equation}
\det K = \prod_{i=1}^{m-1} \det{A_i} \det(A_m - (-1)^m B_m \prod_{i=1}^{m-1} A_{i}^{-1} B_i)
\label{eq:c3det}
\end{equation}

The following sign choice satisfies the parity rules (\ref{eq:signs}) and gives the canonical signs for the terms appearing in the characteristic polynomial (\ref{eq:det}):

\begin{equation}
a_{ij} \rightarrow 1,\, b_{ij} \rightarrow -1,\, c_{ij} \rightarrow -1
\label{eq:c3subs}
\end{equation}
Applying this substitution to $K$ and taking the determinant is a computationally efficient way to compute the multiplicities to high rank of the orbifold group.  
Noting that after the substitution (\ref{eq:c3subs}) the matrices $A, B$ commute, the determinant (\ref{eq:c3det}) reduces to

\begin{equation}
P(z,w) = \det(A^m - (-1)^m B^m w)
\end{equation}
which is easily evaluated.  For a visually appealing example, see figure \ref{fig:c10.30}.

As we discuss in the next section, if the edge weights are kept distinct when taking the determinant, one may obtain the phases and multiplicities of toric subdiagrams by Higgsing (removing points from the diagram).

\subsection{Orbifolds of the conifold}
\label{sec:conifold}

\begin{equation}
K(z,w) = \left(\begin{array}{ccccc}
A_{1} z &0    &\cdots& 0     & B_{1}\\
B_{2} &A_{2}&     0& \cdots& 0 \\
0     &B_{3}&A_{3} & 0     & 0 \\
\ddots&     &\ddots&\ddots&\ddots\\
0 & \cdots & 0 & B_{m} & A_{m}\\
\end{array} \right)
\end{equation}

where

\begin{eqnarray}
A_{i}&=&\left(\begin{array}{cccccc}
c_{i1} & b_{i1} w & 0 & \cdots & \cdots & 0 \\
0 & c_{i 2} & b_{i 2} & 0 & \cdots & 0 \\
\vdots & \ddots & \ddots & \ddots & c_{i (n-1)} & b_{i(n-1)} \\
b_{i n} & 0 &\cdots & \cdots&0 &c_{i n}
\end{array}\right)\nonumber\\
B_{i}&=&\left(\begin{array}{cccccc}
d_{i1} & a_{i1} w & 0 & \cdots & \cdots & 0 \\
0 & d_{i 2} & a_{i 2} & 0 & \cdots & 0 \\
\vdots & \ddots & \ddots & \ddots & d_{i (n-1)} & a_{i(n-1)} \\
a_{i n} & 0 &\cdots & \cdots&0 &d_{i n}
\end{array}\right)
\end{eqnarray}

An appropriate sign choice is

\begin{equation}
a_{ij} \rightarrow -1,\, b_{ij} \rightarrow -1,\, c_{ij} \rightarrow -1,\, d_{ij} \rightarrow 1
\label{eq:consubs}
\end{equation}
As in the previous section, the determinant of the block matrix can be simplified with this sign choice, giving

\begin{equation}
P(z,w) = \det(A^m - (-1)^m B^m z)
\end{equation}

\section{Removing points from the toric diagram (partial resolution)}
\label{partial_resolution}

Given a toric diagram, one may obtain other toric diagrams as subdiagrams by removing points from the corners (this preserves convexity of the polygon).  Given the interpretation of $P$ as counting of paths on the graph, in order to remove a monomial from $P$ we must cut all paths in the graph with the specified height changes. This can be done by removing certain edges from the graph, i.e.~setting their weights to zero.  There is often a choice of which edge(s) to cut to implement removal of a given set of points from the toric diagram, and one finds that the set of multiplicities of the points in the resulting diagram enumerate the possible phases of the toric space that may be obtained by Higgsing, as studied in \cite{Feng:2002zw}.


Certain graphs, such as the hexagonal and square graphs, are universal in the sense that any other toric diagram may be obtained from them by taking a suitably large orbifold by $\Z_m \times \Z_n$ and then removing points from (Higgsing) the resulting toric diagram.  This provides a systematic computational procedure for investigating the phase structure of the associated gauge theories.  In practise, in order to simplify the calculation it is desirable to choose an minimal embedding; i.e.~an embedding into a larger graph with a minimal number of additional points.  The following algorithm may be used:

The adjacency matrix of the embedding graph is written with all edge weights distinct and its' determinant is computed.  Then evaluation of the phases of the toric diagram may be achieved by solving the set of simultaneous equations specifying that the coefficients in the determinant of the points to be removed from the toric diagram all vanish, while the coefficients of the remaining points do not.  Once the set of solutions has been found, the remaining edge weights may be set to a suitable choice of $\pm 1$ to recover the set of phases and multiplicities of the target diagram.

A refined version of this algorithm is as follows:

\begin{enumerate}
\item{Compute the characteristic polynomial of the ambient toric diagram with all edge weights parameterized distinctly (as in Figure \ref{fig:edgeweights}).}
\item{Identify the set of vertices to be removed from the ambient toric diagram, ordered so that the polygon remains convex after each removal.  These are by assumption corner points of the toric diagram at each stage of the removal.}
\item{Take the complement of this set, i.e.~identify the points in the diagram that we do not wish to remove.  For each point (corresponding to a monomial in the determinant), identify the edge weights in the coefficient of this monomial that are common to all terms, i.e.~which factor out.  We cannot set any of these variables to zero, or an unwanted extra point will be removed from the toric diagram.\label{step3}}
\item{For each vertex in the list to be removed, identify the edge weights in the coefficient of the determinant that are common to all terms, i.e.~which factor out.  Discard any edge weights that are on the list computed in step \ref{step3}.\label{step4}}
\item{For each variable in this list, set it to zero to remove the point from the toric diagram, and recurse to step \ref{step4} with the new set of corners of the sub-diagram.}
\item{If all possible cuts have been made (i.e.~no more variables remain to be zeroed), and we have removed all of the points in the target list, we have found a solution.  The edges that were cut to obtain this graph are known, and the multiplicities of this phase may be read off by evaluating the determinant against the variables set to zero and the canonical sign assignment of the initial graph (e.g. (\ref{eq:c3subs}) for $\C^3$).}
\end{enumerate}
The limitations of this algorithm are discussed in the following section.  One advantage of the algorithm compared to methods discussed in previous sections is that it operates on the graph, not the determinant (which contains much less information).  Thus, solutions come not just with the toric multiplicities but the subgraph of the initial graph which produces them.

Note that sometimes it is possible to obtain the desired shape of the toric diagram such that one or more of the corner points have multiplicity greater than one.  This typically corresponds to an inconsistent quiver gauge theory.  However, in all examples we have encountered, it is always possible to perform additional edge cuts to reduce this multiplicity to 1 (this is taken care of by the form of the algorithm presented above).

Using this method we are able to recover all previously known results on multiplicities of toric diagrams (low-rank orbifolds of $\C^3$ and the conifold; cones over $F_0, dP_{0,1,2,3}$, the toric spaces $Y^{p,q}$ and $X^{p,q}$, and some other miscellaneous examples), as well as generating many new examples which have not previously been studied.  Some simple new examples are presented here:

\begin{itemize}
\item{The orbifold $\C^3/(\Z_2 \times \Z_4)$ with point $(0,4)$ removed.}

\begin{equation}
\left[\begin{array}{ccc}
1 & -2 & 1 \\
-3 & -9 &\\
3 & -1 & \\
-1 &&
\end{array}\right]
\end{equation}

\item{The orbifold $\C^3/(\Z_4 \times \Z_4)$ with points $(0,4), (4,0)$ removed.  There are 6 phases:}

\begin{eqnarray}
\left[\begin{array}{cccc}
          1& -3& 3& -1\\
          -3& -65& -43& -1\\
          3& -51& 2& \\
          -1& -1& & 
\end{array}\right]\hspace{0.2in}
\left[\begin{array}{cccc}
          1& -3& 3& -1\\
          -3& -71& -53& -1\\
          3& -45& 2& \\
          -1& -1& & 
\end{array}\right]\hspace{0.2in}
\left[\begin{array}{cccc}
          1& -3& 3& -1\\
          -3& -81& -59& -1\\
          3& -59& 2& \\
          -1& -1& &
\end{array}\right]\nonumber\\
\left[\begin{array}{cccc}
          1& -3& 3& -1\\
          -3& -65& -51& -1\\
          3& -43& 2& \\
          -1& -1& & 
\end{array}\right]\hspace{0.2in}
\left[\begin{array}{cccc}
          1& -3& 3& -1\\
          -3& -71& -45& -1\\
          3& -53& 2& \\ 
          -1& -1& & 
\end{array}\right]\hspace{0.2in}
\left[\begin{array}{cccc}
          1& -3& 3& -1\\
          -3& -73& -43& -1\\
          3& -43& 2& \\
          -1& -1& & 
\end{array}\right]
\end{eqnarray}
\end{itemize}
It is interesting to observe the symmetry in the above set of phases; two pairs of phases are interchanged by reflection, and the other two phases are symmetric.  This structure should also be observed in the phases of the dual quiver gauge theory (the phases of the toric space $Y^{6,0}$ listed in the appendix exhibit a similar structure).

Other toric diagrams may be computed similarly (Mathematica code available upon request).

\section{Discussion}

We have proposed the existence of a duality between quiver gauge theories and dimer models, and developed dimer model techniques for enumerating the toric phases of the gauge theories, and for computing their linear sigma model field multiplicities.   We obtained general formul\ae\ for the multiplicities of arbitrary orbifold theories, and discussed how to obtain the multiplicities of non-orbifold phases by Higgsing a larger toric diagram.  These techniques reproduce all known results from existing quiver gauge theories, and give many predictions for quiver theories not previously examined.

As currently understood, the Higgsing algorithm contains two computational limitations:

\begin{enumerate}
\item{The need to take the determinant of the orbifold adjacency matrix with all edge weights distinct.  Even making use of the block structure of the orbifold matrix as in section \ref{sec:block}, it becomes prohibitively difficult to calculate the determinant for higher orbifold rank (e.g. beyond $\C^3/(\Z_6 \times\Z_6)$ or so, i.e.~a $36 \times 36$ matrix with $3\times36=108$ distinct edge weights).  With a better understanding of the relationship between a given toric diagram and the properties of the dimer model(s) that generate it, it may be possible to identify the edges to be removed without needing to solve the constraint equations, in which case it would not be necessary to use distinct edge weights during the computation.  As figure \ref{fig:c10.30} suggests, this would allow the procedure to be extended to ambient spaces with much higher orbifold ranks.)}
\item{The brute-force enumeration of solutions, which does not account for the large amount of symmetry among the set of solutions.  For example, a set of removed edges may often be translated around the graph without changing the characteristic polynomial.  In practice this means that the running time of the algorithm grows quickly with the number of points to be removed from the diagram, because of a large overcounting of solutions.  It would be useful to precisely characterize and remove the redundancy among solutions.

Typically one finds that the algorithm produces many of the phases very quickly, with the exhaustive search for all phases taking longer.  A shortcut would be to take these first solutions and use Seiberg duality to fill out the set of phases.  The interpretation of Seiberg duality in dimer language will be discussed  in \cite{Kennaway:2005b}.}
\end{enumerate}

The reason for the correspondence between quiver gauge theories and dimer models, and the underlying string theoretical interpretation will be discussed in forthcoming work \cite{Kennaway:2005b}.

{\bf Acknowledgements:} 
We would like to thank Sergio Benvenuti, Alan Dunn, Sebastian Franco, Marco Gualtieri, Yang-Hui He, Sheldon Katz, Pavlos Kazakopoulos, David Vegh and Brian Wecht for many enlightening discussions.
A.~H.~would like to thank the Fields Institute for Mathematical Sciences and the Physics Department of the University of Toronto for their hospitality while this work was initiated, and conversely K.~K.~thanks the Center for Theoretical Physics, MIT, for theirs while it was completed.
A.~H.~would like to thank the department of physics in Amsterdam University for their hospitality while final stages of this paper were completed.
The research of  A.~H.~was supported in part by the CTP and LNS of
MIT and the U.~S.~Department of Energy under cooperative research agreement $\#$ DE--FC02--
94ER40818, and by BSF, an American--Israeli Bi--National Science Foundation. A.~H.~is also
indebted to a DOE OJI Award.  K.~K.~is supported by NSERC.

\newpage
\EPSFIGURE[t]{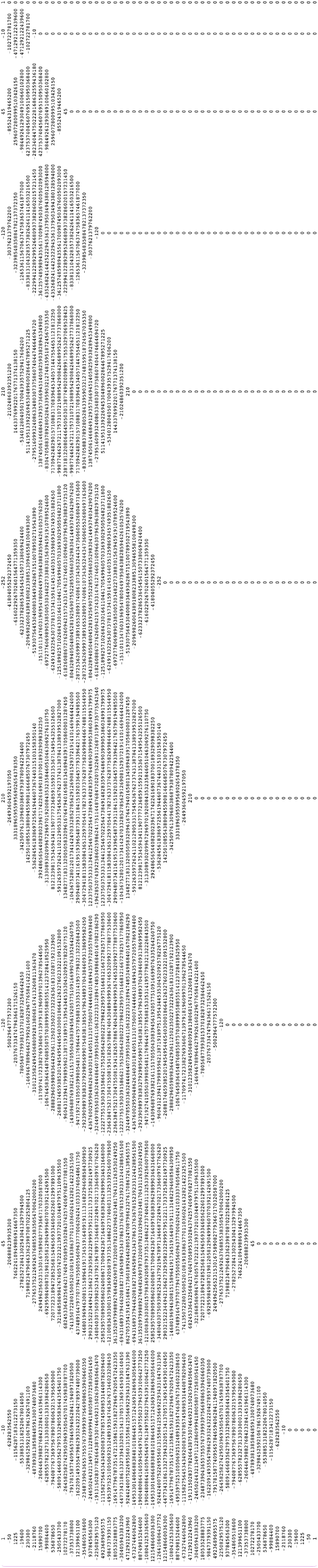,height=8in,width=4in}{Multiplicities for the orbifold $\C^3/(\Z_{10} \times \Z_{50}).$\label{fig:c10.30}}

\begin{appendix}
\section{Toric phases of $Y^{6,0}$}
\label{app:60}

The following are the multiplicities of 17 of the 18 phases \cite{Brian} of the toric space $Y^{6,0}$ (suppressing the minus signs from (\ref{eq:det})), and the edges in the graph of the $\Z_2 \times \Z_6$ orbifold of the conifold (using the labeling of figure \ref{fig:edgeweights} and the block matrix described in section \ref{sec:conifold}) that may be cut to obtain this phase. Many other equivalent choices of edges are possible.  The ``missing'' 18th phase was identified in \cite{Franco:2005fd} and shown to have identical toric multiplicities to one of the remaining 17.

\begin{eqnarray}
\left[\begin{array}{ccccccc}
&&&1&&&\\
1&14&61&102&65&15&1\\
&&&1&&&\\
\end{array}\right]&\hspace{0.5in}& d11, d21, d31, a41, a51, a61, c12, b22, c42, b32, c62, b52 \nonumber\\
\left[\begin{array}{ccccccc}
&&&1&&&\\
1&15&65&102&61&14&1\\
&&&1&&&\\
\end{array}\right]&\hspace{0.5in}&d11, d21, d31, a41, a51, a61, c22, b12, c32, b42, c62, b52\nonumber\\
\left[\begin{array}{ccccccc}
&&&1&&&\\
1&14&63&106&67&15&1\\
&&&1&&&\\
\end{array}\right]&\hspace{0.5in}&d11, d21, d41, a31, a51, a61, c12, b22, c42, b32, c62, b52\nonumber\\
\left[\begin{array}{ccccccc}
&&&1&&&\\
1&15&67&106&63&14&1\\
&&&1&&&\\
\end{array}\right]&\hspace{0.5in}&d11, d21, d41, a31, a51, a61, c22, b12, c32, b42, c62, b52\nonumber\\
\left[\begin{array}{ccccccc}
&&&1&&&\\
1&12&48&76&48&12&1\\
&&&1&&&\\
\end{array}\right]&\hspace{0.5in}&d11, d21, d31, a41, a51, a61, c32, b12, c42, b22, c52, b62\nonumber\\
\left[\begin{array}{ccccccc}
&&&1&&&\\
1&13&54&86&54&13&1\\
&&&1&&&\\
\end{array}\right]&\hspace{0.5in}&d11, d21, d31, a41, a51, a61, c12, b22, c32, b52, c42, b62\nonumber\\
\left[\begin{array}{ccccccc}
&&&1&&&\\
1&13&54&87&54&13&1\\
&&&1&&&\\
\end{array}\right]&\hspace{0.5in}&d11, d21, d31, a41, a51, a61, c12, b22, c32, b42, c52, b62\nonumber
\end{eqnarray}
\begin{eqnarray}
\left[\begin{array}{ccccccc}
&&&1&&&\\
1&13&55&89&55&13&1\\
&&&1&&&\\
\end{array}\right]&\hspace{0.5in}&d11, d21, d31, a41, a51, a61, c22, b12, c42, b32, c52, b62\nonumber\\
\left[\begin{array}{ccccccc}
&&&1&&&\\
1&14&62&104&62&14&1\\
&&&1&&&\\
\end{array}\right]&\hspace{0.5in}&d11, d21, d31, a41, a51, a61, c22, b12, c42, b32, c62, b52\nonumber\\
\left[\begin{array}{ccccccc}
&&&1&&&\\
1&14&63&102&63&14&1\\
&&&1&&&\\
\end{array}\right]&\hspace{0.5in}&d11, d21, d41, a31, a51, a61, c32, b12, c42, b22, c62, b52\nonumber\\
\left[\begin{array}{ccccccc}
&&&1&&&\\
1&14&63&103&63&14&1\\
&&&1&&&\\
\end{array}\right]&\hspace{0.5in}&d11, d21, d41, a31, a51, a61, c12, b22, c32, b52, c42, b62\nonumber\\
\left[\begin{array}{ccccccc}
&&&1&&&\\
1&14&64&109&64&14&1\\
&&&1&&&\\
\end{array}\right]&\hspace{0.5in}&d11, d21, d41, a31, a51, a61, c12, b22, c32, b42, c52, b62\nonumber\\
\left[\begin{array}{ccccccc}
&&&1&&&\\
1&15&63&100&63&15&1\\
&&&1&&&\\
\end{array}\right]&\hspace{0.5in}&d11, d21, d31, a41, a51, a61, c12, b22, c52, b32, c62, b42\nonumber\\
\left[\begin{array}{ccccccc}
&&&1&&&\\
1&15&65&105&65&15&1\\
&&&1&&&\\
\end{array}\right]&\hspace{0.5in}&d11, d21, d31, a41, a51, a61, c12, b22, c32, b42, c62, b52\nonumber\\
\left[\begin{array}{ccccccc}
&&&1&&&\\
1&15&75&140&75&15&1\\
&&&1&&&\\
\end{array}\right]&\hspace{0.5in}&d11, d31, d51, a21, a41, a61, c22, b12, c42, b32, c62, b52\nonumber\\
\left[\begin{array}{ccccccc}
&&&1&&&\\
1&16&78&132&78&16&1\\
&&&1&&&\\
\end{array}\right]&\hspace{0.5in}&d11, d21, d41, a31, a51, a61, c12, b22, c32, b52, c42, b62\nonumber\\
\left[\begin{array}{ccccccc}
&&&1&&&\\
1&18&79&126&79&18&1\\
&&&1&&&\\
\end{array}\right]&\hspace{0.5in}&d11, d21, d31, a41, a51, a61, c12, b22, c32, b42, c62, b52\nonumber\\
\end{eqnarray}
\end{appendix}


\bibliographystyle{JHEP}
\bibliography{kk}

\end{document}